\documentclass[twocolumn,amsmath,amssymb,aps]{revtex4}

\usepackage{graphicx}
\usepackage{dcolumn}
\usepackage{bm}
\usepackage{float}
\begin{document}

\preprint{PREPRINT}

\title{Measurements of Accelerations of Large Neutrally-buoyant Particles in Intense Turbulence}

\author{Rachel D. Brown$^a$}
\author{Zellman Warhaft$^b$}
\author{Greg A. Voth$^a$}\email{gvoth@wesleyan.edu}

\affiliation{$^a$Department of Physics, Wesleyan University, Middletown,
CT  06459, U.S.A.}
\homepage{http://gvoth.web.wesleyan.edu/lab.htm}

\affiliation{$^b$Sibley School of Mechanical and Aerospace Engineering, Cornell University, Ithaca,
NY 14853, U.S.A.}

\date{\today}

\begin{abstract}
We measure acceleration statistics of neutrally buoyant spherical particles with diameter $0.4 < d/\eta <27$ in intense turbulence ($400< R_\lambda < 815$). High speed cameras image polystyrene tracer particles in a flow between counter-rotating disks. The measurements of acceleration variance, $\langle a^2 \rangle$, clearly resolve the transition from the tracer like behavior of small particles to the much smaller accelerations of large particles.   For $d > 5\eta$, $\langle a^2 \rangle$ decreases with diameter as $d^{-2/3}$ in agreement with inertial range scaling arguments. A model relating $\langle a^2 \rangle$ to the pressure structure functions  matches the transition from small to large particle behavior if the particles respond to pressure differences over $(1.7 \pm 0.3) d$.  A model relating $\langle a^2 \rangle$ to the fluid acceleration averaged over the particle diameter predicts the transition with no free parameters, but does not show clean inertial range scaling in the size range studied. Consistent with earlier work, we find that the scaled acceleration probability density function shows very little dependence on particle size.
\end{abstract}

\maketitle

The study of accelerations of turbulent fluids and of particles entrained in turbulent fluids has produced many recent insights~\cite{toschi:2009}.  Even excluding differences between turbulent flows, the parameter space to describe particle motion contains a wide range of possibilities spanning both particle size and particle density.  The most intensely studied region of this parameter space has been small particles that are heavier than the fluid~\cite{squires:1991, ayyalasomayajula:2006}.  The motion of small heavy particles is relevant to many natural systems including water droplets in clouds and sedimentation.  This problem is also attractive because it is accessible numerically by modeling the particles as points interacting with the fluid through a drag law~\cite{bec:2006a,calzavarini:2008PRL}.  Significant study has also focused on the region of the particle parameter space in which the particles have positive buoyancy, such as air bubbles in water~\cite{Mazzitelli:2004, volk:2008}.  Our interest in this paper is in the accelerations of neutrally buoyant particles that vary in diameter from tracer particles ($d=0.4\eta$) to much larger than the Kolmogorov length ($d=27\eta$).  A number of  applications, including many marine organisms, inhabit the large neutrally buoyant domain.  Additionally, tracer particles used in liquid phase fluid measurements are often neutrally buoyant.


Several previous experiments have addressed the problem of large neutrally buoyant particles in turbulent flow. Voth \textit{et al.}(2002) measured size dependence of particle accelerations in order to confirm tracer particle behavior in their experiment.  They have relatively few data points; however, they suggested that the accelerations of these large particles might be modeled by terminating the cascade at the particle diameter, $d$.  The particle acceleration then scales with the fluid acceleration at the scale of the particle diameter, which implies $\langle a^2_{\mathrm{particle}}\rangle \propto d^{-2/3}$ in the Kolmogorov inertial range.   Qureshi \textit{et al.}\cite{qureshi:2007} measured accelerations of neutrally buoyant helium bubbles in a wind tunnel.  Their particles ranged from 7$\eta$ to 25$\eta=0.1L$ which at their relatively low Reynolds number ($R_\lambda = 160$), spans from the dissipation scales to the injection scales.   But they are unable to observe the transition from dissipation to inertial range.    They highlight the fact that the particle acceleration variance can be connected to the pressure structure function evaluated at the diameter of the particle, $\langle a^2_{\mathrm{particle}}\rangle \propto \langle \Delta P^2(d) \rangle/d^2$ which gives the same $d^{-2/3}$ scaling in the inertial range for pressure structure functions that have $r^{4/3}$ inertial range scaling.

Numerical simulations of this problem are difficult because a complete model must resolve the turbulent scales and the boundary layer around individual particles~\cite{Cate:2004,zhang:2005}.  Calzavarini \textit{et al.} overcome the severe Reynolds number limitations of full simulations by using effective equations for particle motion that utilize phenomenological Fax\'{e}n corrections~\cite{calzavarini:2009}.

In this letter we present experimental measurements of the acceleration of neutrally buoyant particles at large Reynolds numbers and a range of particle size that spans from dissipative lengthscales into the inertial range.   These measurements are the first to clearly resolve the transition from tracer particle to large particle behavior and allow a stringent test of various models that have been proposed to quantify the accelerations of large particles.
\begin{center}
\begin{table*}
\begin{tabular*}{0.9\textwidth}%
     {@{\extracolsep{\fill}}|c|c|c|c|c|c|c|c|c|c|c|}
\hline
Disk     &$\bm{\nu}$&$\bm{R_\lambda}$&&&$\bm{\epsilon}$&$\bm{\eta}$&$\bm{\tau_\eta}$&&Imaging &\\
Frequency (f)&($\times 10^6$)&$=(15 \tilde{u} L/\nu)^{1/2}$&\raisebox{1.5ex}{$\bm{\tilde{u}}$}&\raisebox{1.5ex}{$\bm{L}$}&$=\tilde{u}^3/L$&$=(\nu^3 / \epsilon)^{1/4}$&$=(\nu/\epsilon)^{1/2}$&\raisebox{1.5ex}{$\bm{N_f}$}&Volume&\raisebox{1.5ex}{$\bm{\Delta x}$}\\

\hline
Hz&m$^2$/s&&m/s&mm&m$^2$/s$^3$&$\mu$m&ms&$\mathrm{Frames}/\tau_\eta$&$\eta^3$&$\mu$m/pix\\
\hline
5.25&1.00&813&0.62&71&3.41&23.3&0.54&11&$430^3$&41.1\\
5.25&1.29&717&0.62&71&3.41&28.2&0.62&12&$355^3$&41.1\\
1.6&1.00&449&0.19&71&0.10&56.9&3.23&64&$175^3$&41.1\\
1.6&1.29&396&0.19&71&0.10&68.7&3.66&73&$145^3$&41.1\\
\hline
\end{tabular*}
\caption{\label{tab:par} Table of flow parameters: $f$, frequency of the rotating disks; $\nu$, kinematic viscosity of the fluid; $R_\lambda$, Taylor Reynolds number; $\tilde{u}$, rms velocity of the flow; $L$, energy input length scale; $\eta$, Kolmogorov length scale; $\tau_\eta$, Kolmogorov time scale; $N_f$, number of frames in each Kolmogorov time; imaging volume;  and $\Delta x$, the distance in the flow corresponding to one pixel. }
\end{table*}
\end{center}
Lagrangian particle tracking measurements were carried out in the von Karman flow between counter rotating disks described in  ~\cite{voth:2002}. The relevant parameters of the flow, including $R_\lambda$ and the Kolmogorov microscales, are shown in Table~\ref{tab:par}.  Three Phantom v7 cameras operating at 20,000 frames per second and $256 \times 256$ pixels per frame were arranged in the central horizontal plane and focused on a 1 cm$^3$ region at the center of the flow.  A frequency-doubled, pulsed Nd:YAG laser with 50W average power was used for illumination. Two beams were required to allow for forward scattering in all three cameras. Vertical polarization minimizes secondary reflections from the particles~\cite{gvthesis}.

All particles used were polystyrene spheres $(\rho=1.05$ g/cm$^3$).  Measurements were made with the flow seeded with both mono-dispersed particles from Duke Scientific with known diameters ($d$=26, 55, 134, 222, 300, 400 $\mu$m)  and with poly-dispersed `grinding media' with diameters in the range 600-990 $\mu$m obtained from Norstone, Inc.  We use a reflection pair method to determine the diameter of the large particles from the images of each trajectory.  The central camera records two reflections from each particle in the imaging volume because of the two beams.  For large particles ($d >$ 100 $\mu$m) the two reflections can be resolved, allowing measurements of particle size in every frame. Excluding particles whose diameter changes over time minimizes error due to the small fraction of non-spherical particles in the grinding media.

An $8 \% $ NaCl solution by mass $(\nu=1.29 \times 10^6\mathrm{ m^2/s},  \rho=1.048 \mathrm{g/cm^3})$ is used in order to density match particles with diameters greater than 100 microns, but deionized water is used for particles with diameters of 26 and 55 microns. The effect of a $5\%$ density mismatch on particle accelerations is nearly negligible for all our particle sizes~\cite{voth:2002}; however keeping the large particles suspended in the fluid requires careful density matching.

\begin{figure}[b!]
 \begin{center}
\includegraphics[width=2.0in]{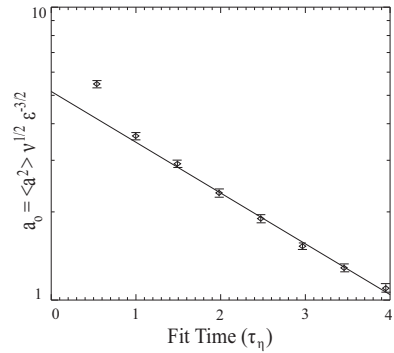}
\caption{\label{fig:accvarfit} An example extrapolation of acceleration variance back to zero fit time.  Acceleration variance measurements for single reflection measurements of the $300 \mu $m ($d/ \eta=10.64$) particles taken at $R_{\lambda} =717$ are plotted at various fit times, along with the statistical error in these values.}
 \end{center}
\end{figure}

We use the data acquisition system used in~\cite{Ouellette:2006}. Images are acquired in one second sequences, which take 3-5 minutes to download.  Particle positions extracted from the images are stored and later tracked.  For each particle size, we obtain on average 65 million particle positions from 40-500 sequences. Accelerations are determined with a quadratic fit to the particle trajectories.  Figure~\ref{fig:accvarfit} shows the measured acceleration variance for one data set as a function of fit time.  For the highly intermittent accelerations of particles in turbulence, it is very difficult to obtain fits that are not affected either by measurement errors or by smoothing of the trajectories.  We use an established method~\cite{voth:2002} to extrapolate back to zero fit time in order to determine the acceleration variance.

All results reported are for analysis of two dimensional trajectories.  With three cameras, it is possible to determine 3D trajectories.
However, we encountered difficulties with stereomatching from images of particles that fill a large fraction of the field of view and so the 3D tracks are extremely fractured.  As a result, we find much better statistical convergence for the 2D trajectories. In this letter, we focus on the axial component of acceleration (parallel to the axis of rotation of the disks in our flow).  The transverse component shows similar results with larger measurement uncertainties due to the illumination geometry.  In the Reynolds number range studied, the transverse acceleration variance of fluid particles in this flow is 10\% to 20\% larger than the axial variance~\cite{voth:2002}.

\begin{figure}[tb!]
 \begin{center}
\includegraphics[width=2.7in]{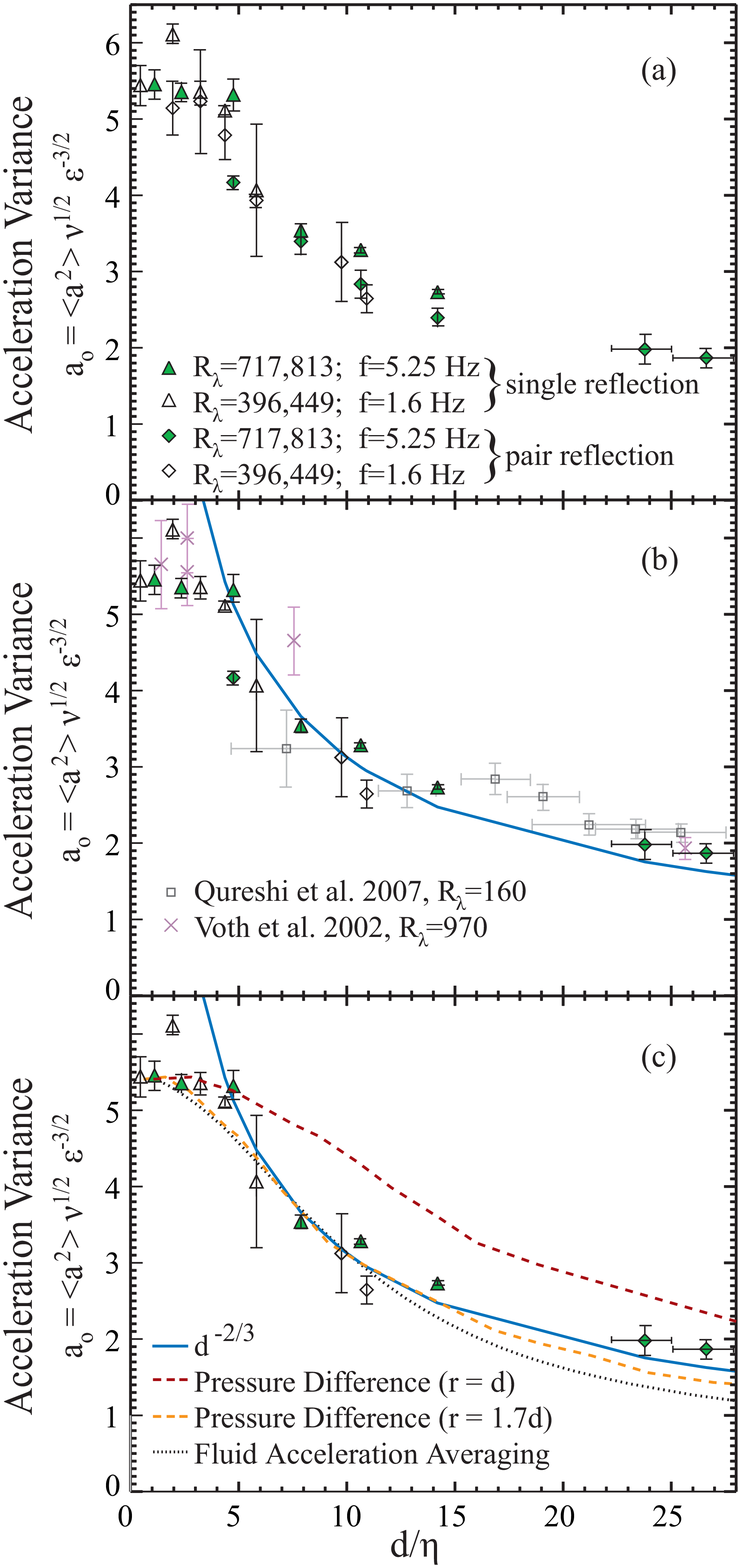}
\caption{\label{fig:accvarcurves} (color online) (a) Acceleration variance as a function of particle diameter.  Unfilled symbols represent data taken at the lower disk frequency and hence lower Reynolds numbers ($R_{\lambda}=396,449$). Filled symbols represent data taken at the higher disk frequency and hence higher Reynolds numbers ($R_{\lambda}=717,813$).  Triangles are single reflection measurements.   Diamonds are reflection pair measurements. (b)  Filled and unfilled symbols are the same as in A, but pair measurements are not shown for data sets where single reflection measurements are available.  X's are  measurements from~{\protect \cite{voth:2002}}.  Grey squares from~{\protect \cite{qureshi:2007}}. The solid curve is the the inertial scaling derived from the cascade truncation model.
(c) Symbols are the same as in A. The solid curve is the $d^{-2/3}$ inertial range scaling.   The  dashed curves show the predictions of the Pressure Difference model.  The red dashed line uses a separation equal to the particle diameter. The dashed orange line uses the separation which best fits fit to our data ($1.7d$).  The dotted black curve is the prediction of the Fluid Acceleration Averaging model.}
 \end{center}
\end{figure}

Figure~\ref{fig:accvarcurves}(a) shows acceleration variance measurements as a function of particle diameter.  The acceleration is normalized by Kolmogorov variables to give $a_0$.    For small particle diameters ($d/\eta \le 5$) the normalized acceleration variance is nearly constant.  At larger diameters the normalized acceleration variance falls off implying that these large  particles do not simply trace the flow.  There is no detectable dependence on Reynolds number.  However, at the lower Reynolds number, the largest particles studied are only 10.9$\eta$.   For fluid particles, $a_0$ is found to vary by less than 17\% over the Reynolds number range we study~\cite{voth:2002}.
Data is presented for two methods of determining particle positions. The first method treats each reflection observed in an image as a particle. This method works well for mono-dispersed particles, $d \le 400 \mu m$, and we can use data from all three cameras to improve statistical convergence.    The second method was developed to determine the size of polydisperse particles from the images.  Here we use only the central camera, and require two reflections to be observed that move parallel to each other.  This analysis can be done for  particles with diameter $d \ge 134 \mu$m.  Data points are plotted for both of these methods to show that no systematic error is introduced when we use the reflection pair method. Error bars in Figure~\ref{fig:accvarcurves} represent statistical uncertainties. Systematic errors associated with determination of the energy dissipation rate and extrapolation also exist~\cite{gvthesis}.  Figure~\ref{fig:accvarcurves}(b) includes data from two previous measurements of accelerations of large particles.

Figure~\ref{fig:accvarcurves}(c) shows the measured acceleration variance compared with predictions from three models.  The first of these models is the $d^{-2/3}$ scaling predicted for particles with diameters in the inertial range~\cite{voth:2002}.  Our data shows good agreement with the $d^{-2/3}$ scaling for $d>5\eta$.

A second model is suggested by the connection between the particle acceleration variance and the pressure structure function~\cite{qureshi:2007}.
In addition to showing $d^{-2/3}$ inertial range scaling, this Pressure Difference model can be extended to predict the acceleration variance for small particle diameters $d < \eta$ using the pressure structure function for separation distances in the dissipation range.
  In a previous experiment in the same apparatus that we use, Xu \textit{et al.} ~\cite{xu:2007} measured the pressure structure function over the range $0.7 \eta <r< 140 \eta$.  We use their data to create the prediction shown as a dashed line in Fig.~\ref{fig:accvarcurves}.
When the separation at which the pressure difference is sampled is equal to the particle diameter, this model predicts larger accelerations than we measure at large particle sizes.   However, this model is a good fit when we allow the separation to be an adjustable parameter and sample the
pressure structure functions at $(1.7\pm0.3)d$. This raises questions about the mechanism which would cause the particle's acceleration to be determined by lengthscales larger than its diameter.

The third model assumes that the particle acceleration is equal to the spatial average of the fluid acceleration over the volume of a particle.  This model closely matches the model used by Calzavarini ~\textit{et al}~\cite{calzavarini:2009}.  Their model differs by a drag term that contributes only 1\% to the neutrally buoyant particle acceleration.

In this Fluid Acceleration Averaging model, the particle acceleration variance is given by
\begin{equation}
\langle a_{\mathrm{particle}}{}^2 \rangle =
\int_v \int_{v'} \langle a_f(\mathbf{r})a_f(\mathbf{r'}) \rangle d^3\mathbf{r}d^3\mathbf{r'}.
\end{equation}
Integrating and using the isotropic expression, $R_{ij}=R_{NN} \delta_{ij} + (R_{LL}-R_{NN}) r_i r_j / r^2 $ for the acceleration correlation function yields
\begin{equation}
\frac{\langle a_{\mathrm{particle}}{}^2 \rangle} {\langle a_{f}{}^2 \rangle}
=\frac{8}{d^3} \int_0^{d/2} \left(2R_{NN}(r) + R_{LL}(r)\right) r^2 dr .
\label{eq:accav}
\end{equation}
Xu \textit{et al.}~\cite{xu:2007} have measured the acceleration correlation functions, $R_{LL}(r)$ and $R_{NN}(r)$, in the flow that we study.  Integrating their measured functions in Eq.~\ref{eq:accav} produces the dotted curve shown in  Fig.~\ref{fig:accvarcurves}(c). This model gives a very good fit to our data both at small scales and in the transition to inertial range scaling.  No free parameters are needed to adjust the length scale in the model to the measurements as is required for the pressure structure function model.
However, this third model predicts accelerations that are somewhat too small for large particles sizes.  It also does not display $d^{-2/3}$ scaling in the range of particle sizes we studied.  The acceleration correlation functions do not show inertial range scaling until $d>40 \eta$~\cite{xu:2007}, and so inertial range scaling for this model is not expected in the range $d<25 \eta$ that we study.  Although the data seems to match the $d^{-2/3}$ scaling quite well, the lack of scaling in the acceleration correlation functions suggests the possibility that the observed $d^{-2/3}$ scaling for small particles sizes is only approximate and that much larger particles sizes (along with high Reynolds numbers) are needed before rigorous $d^{-2/3}$ scaling will exist.
\begin{figure}[t!]
 \begin{center}
\includegraphics[width=2.3in]{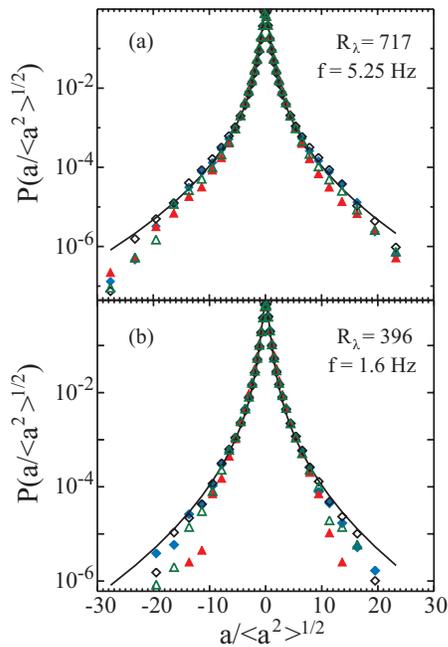}
\caption{\label{fig:pdf_freq525}Probability density functions of acceleration at Taylor Reynolds numbers (a) 717 and (b) 396.  Particle sizes in $d/ \eta$ represented by each symbol are as follows: black unfilled diamonds, (a) $1.95$ and (b) $4.75$; blue filled diamonds, (a) $3.23$ and (b) $7.87$; green unfilled triangles,(a) $4.37$ and (b) $10.64$; and red filled triangles, (a) $5.82$ and (b) $14.18$.}
 \end{center}
\end{figure}
Figure ~\ref{fig:pdf_freq525} shows the probability density function (pdf) of particle accelerations normalized by the standard deviation for particle sizes $d<15 \eta$. Within our measurement error, these pdfs agree with the form of the fluid acceleration pdf ~\cite{mordant:2004}.  In the tails of the pdfs, the probability densities for large particles sizes may be slightly below the fluid particle pdf.  This is more pronounced in the data from the large poly-dispersed particles (not shown).  However, in looking carefully at our data we found errors due to particle finding, tracking, and nonspherical or stray particles that seem to affect the rare events by as much as any discrepancy with the fluid acceleration pdf, so we do not draw any conclusions from these slight differences. This particle size independence of the acceleration pdfs has been observed before~\cite{qureshi:2007,voth:2002,Xu:2008}  and remains somewhat puzzling.


These measurements of accelerations of spherical neutrally buoyant particles in intense turbulence clearly resolve the transition from the tracer regime to the large particle regime.  Our measurements of the acceleration variance as a function of particle size show the best agreement yet measured with the inertial range $d^{-2/3}$ scaling.  We present two models that capture the transition from tracer to large particle behavior.  The success of these simple models is perhaps surprising given the complex flow in the boundary layer around large particles in turbulence. The parameter space of this problem is vast, but these measurements seem to put our understanding of one region, large neutrally buoyant particles, on a solid foundation.

\section{\label{sec:acknowledgements}Acknowledgements}
This work was supported by NSF grants DMR-0547712 to Wesleyan University and NSF-0756501 to
Cornell.  We thank Eberhard Bodenschatz, Haitao Xu, Nick Ouellette, Mark Nelkin, Stephanie Neuscamman and Sergey Gerashchenko.

\bibliographystyle{apsrev}

\end{document}